\documentclass[fleqn,usenatbib]{mnras}

\usepackage{newtxtext,newtxmath}

\usepackage[T1]{fontenc}
\usepackage{ae,aecompl}
\usepackage{hyperref}
\usepackage{adjustbox}

\usepackage[usenames]{xcolor}
\usepackage{color, soul}

\usepackage{graphicx}	
\usepackage{amsmath}	

\usepackage{amssymb}	
\usepackage{times}
\usepackage{pifont}
\defcitealias{calcino2023}{Paper~I}

\newcommand{\rr}[1]{{{#1}}}

\title[Kinematic Signatures of Circumbinary Discs II]{Observational Signatures of Circumbinary Discs II: Kinematic Signatures in Velocity Residuals}

\author[J. Calcino et al.]
{Josh Calcino$^{1}$\thanks{Contact e-mail: \href{mailto:jcalcino@mail.tsinghua.edu.cn}{jcalcino@mail.tsinghua.edu.cn}},
Brodie Norfolk$^{2}$,
Daniel J. Price$^{2}$,
Thomas Hilder$^{2}$,
Jessica Speedie$^{3}$, 
\newauthor
Christophe Pinte$^{2,4}$,
Himanshi Garg$^{2}$,
Richard Teague$^{5}$, Cassandra Hall$^{6,7}$, Jochen Stadler$^{8,4}$
\\
$^{1}$Department of Astronomy, Tsinghua University, 30 Shuangqing Rd, 100084 Beĳing, China\\
$^{2}$School of Physics and Astronomy, Monash University, Vic 3800, Australia\\
$^{3}$Department of Physics \& Astronomy, University of Victoria, Victoria, BC, V8P 5C2, Canada\\
$^{4}$Univ. Grenoble Alpes, CNRS, IPAG, F-38000 Grenoble, France\\
$^{5}$Department of Earth, Atmospheric, and Planetary Sciences, Massachusetts Institute of Technology, Cambridge, MA 02139, USA \\
$^{6}$Department of Physics and Astronomy, The University of Georgia, Athens, GA 30602, USA \\
$^{7}$Center for Simulational Physics, The University of Georgia, Athens, GA 30602, USA \\
$^{8}$Laboratoire Lagrange, Université Côte d’Azur, CNRS, Observatoire de la Côte d’Azur, 06304 Nice, France 4 \\
}
\date{Accepted XXX. Received YYY; in original form ZZZ}

\pubyear{2023}

\begin{document}
\label{firstpage}
\pagerange{\pageref{firstpage}--\pageref{lastpage}}
\maketitle

\begin{abstract}
Kinematic studies of protoplanetary discs are a valuable method for uncovering hidden companions. In the first paper of this series, we presented five morphological and kinematic criteria that aid in asserting the binary nature of a protoplanetary disc. 
In this work we study the kinematic signatures of circumbinary discs in the residuals of their velocity maps. We show that Doppler-flips, spiral arms, eccentric gas motion, fast flows inside of the cavity, and vortex-like kinematic signatures are commonly observed. \rr{Unlike in the planetary mass companion case, Doppler-flips in circumbinary discs are not necessarily centred on a companion, and can extend towards the cavity edge. }
We then compare the kinematic signatures in our simulations with observations and see similarities to the Doppler-flip signal in HD~100546 and the vortex-like kinematic signatures in HD~142527. Our analysis also reveals kinematic evidence for binarity in several protoplantary disks typically regarded as circumstellar rather than circumbinary, including AB Aurigae and HD 100546.
\end{abstract}




\section{Introduction}
\rr{A detailed analysis of the gas kinematics in protoplanetary discs has proven to be an invaluable tool for understanding the origin of the substructures often found in them \citep{pinte2018,pinte2019,casassus2019,ddc2020,teague2021,pp7kinematics}.}
\rr{Since the velocity field of the gas is a direct product of the forces acting on it, the gas kinematics are a powerful probe of the phenomena responsible for the observed substructures \citep[e.g.][]{dong2018c, francis2020, pinte2023, bae2023}}
Different physical phenomena can produce similar substructures in the gas and dust distribution. For example, rings and gaps seen in the mm dust continuum can be produced by planets \citep[e.g.][]{dippierro2015}, volatile snow lines \citep[e.g.][]{zhang2015}, the irradiation instability (\citealt{wu2021, kutra2023}; however see \citealt{fuksman2022, Pavlyuchenkov2022}), MHD driven zonal flows and winds \citep[e.g.][]{suriano2018}, plus many more \citep[for a review, see][]{bae2023}. \rr{The gas kinematics break this degeneracy as different models often predict different kinematic signatures \citep{pp7kinematics}.}

\begin{table*}
    \centering
    \begin{adjustbox}{width=0.99\textwidth}
    \begin{tabular}{l|c|c|c|c|c|c|c|c|c|c|c|c}
    \hline
        Ref. & $q$    & $a$ (au) & $e$ & $i$ & $\omega$ & $M_\textrm{disc}$ (M$_\odot$) & $H/R_\textrm{ref}$ & $R_\textrm{ref}$ & $R_\textrm{in}$ (au) & $R_\textrm{out}$ (au) & $\alpha_{\textrm{SS}}$ & $N_\textrm{Orbits}$ \\
        \hline
        Planet (P)                      & $2.5\times 10^{-3}$  & 80  & 0.0 & 0.0 & 0.0 & 0.010  & 0.066 & 100   & 10    & 400 & $5\times 10^{-3}$   &  60 \\
        Multiple Planets (MP)           & $[2.5, 1.25]\times 10^{-3}$  & [75.6, 130]  & 0.0 & 0.0 & 0.0 & 0.010  & 0.066 & 100   & 10    & 400 & $5\times 10^{-3}$   &  60 \\
        Eccentric Planet (EP)           & $2.5\times 10^{-3}$  & 80  & 0.4 & 0.0 & 0.0 & 0.010  & 0.066 & 100   & 100    & 400 & $5\times 10^{-3}$   &  60 \\
        \textbf{No Over-density (NOD)}           & 0.25 & 40  & 0.0 & 0  & 0  & 0.010  & 0.066 &  100  & 63    & 400 & $5\times 10^{-3}$   &  1100 \\
        \textbf{Over-density (OD)}               & 0.2  & 30  & 0.0 & 0  & 0  & 0.005  & 0.05  &  45   & 45    & 120 & $1.5\times 10^{-3}$ &  500 \\
        \textbf{Eccentric Companion (EC)}        & 0.1  & 40  & 0.4 & 0  & 0  & 0.010  & 0.066 & 100   & 90    & 400 & $5\times 10^{-3}$   &  80 \\
        \textbf{Heavy Inclined Companion (HIC)}  & 0.25 & 40  & 0.5 & 30 & 0  & 0.010  & 0.066 & 100   & 90    & 400 & $5\times 10^{-3}$   &  20 \\
        \textbf{Polar Companion (PC)}            & 0.2  & 40  & 0.5 & 90 & 90 & 0.010  & 0.066 & 100   & 90    & 400 & $5\times 10^{-3}$   &  60 \\

        \hline
    \end{tabular}
    \end{adjustbox}
    \caption{A summary of the initial conditions of the models presented in Paper I that are used in the present work. Circumbinary disc models are highlighted in bold font. Note that model OD is taken from Calcino et al. (2019), but the disc parameters have been rescaled.}
    \label{tab:ic}
\end{table*}

Recent works have focused on understanding how forming planets produce velocity perturbations in the gas  \citep{perez2015, perez2018, teague2018, bollati2021}, developing tools to aid in their detection and characterisation \citep{izquierdo2021, terry2021}. One way to examine these perturbations is by subtracting a best-fit Keplerian disc model \citep[e.g.][]{perez2018,eddy}. After doing-so, the location of a planet may be revealed by a change in sign of velocity residual across a narrow azimuthal section of the disk, known as a Doppler-flip \citep{perez2018, pp7kinematics}. One such \rr{Doppler-flip was discovered in} HD~100546 by \cite{perez2020} and attributed to a $\sim 10$ M$_J$ planetary companion. However the signature is also consistent with a more massive companion inside the cavity \citep{norfolk2022}. If channel maps are inspected, a perturbation in the iso-velocity curve, known as a ``velocity kink'', may reveal the location of the planet \citep[][]{perez2015, pinte2018}.

Although there are several convincing \rr{indrect} detections of planets using the gas kinematics \citep[e.g.][]{pinte2018, pinte2019, izquierdo2023}, observations also reveal that disc kinematics can be complex \citep{pinte2020, huang2021}. 
Planets are not the only potential source of perturbations in protoplanetary discs. 
However, not much attention has been paid to the dynamical effects of other sources. 
The most well studied is gravitational instability (GI) \citep{hall2020,Longarini2021, veronesi2021, terry2021,lodato2023}, but the conditions for GI are only present in a handful of the most well-studied discs \citep{paneque2021, veronesi2021,lodato2023}. Expanding our understanding of the dynamical effects that other disc substructure producing mechanisms have on the velocity field of the disc can identify or rule out each mechanism. 
For example, dust horseshoes could be vortices induced either in planet-hosting \citep{zhu2015} 
or circumbinary discs \citep{rabago2023}, or they could be over-dense lumps in eccentric discs that lack vortical motion \citep{ragusa2017, calcino2019, ragusa2020}. 
Continuum observations alone may not be enough to distinguish between these scenarios.
It has been found that the anti-cyclonic motion inside of a vortex can be detected in the velocity residuals, 
and appears as a sign change in the velocity residuals across the major axis of the vortex \citep{huang2018, robert2020}. 
\cite{boehler2021} claimed a potential detection of such a signature in HD~142527.
However it is known that this is a binary star system \citep{biller2012, nowak2024}, and it is unclear how the central binary imprints its kinematic signatures on the cavity.

In \citet[][hereafter Paper I]{calcino2023} we showcased some of the kinematic features expected 
to arise in circumbinary discs and developed five criteria to infer binarity. We found three morphological 
features commonly associated with circumbinary discs: i) a gas depleted central cavity; ii) spiral arms inside or outside of the cavity;
and iii) velocity kinks close to the central cavity in the channel maps.
We also devised two kinematic criteria which quantify the degree to which the projected velocity field is perturbed.
Together these morphological and kinematic criteria 
can be used to distinguish circumbinary discs from circumstellar discs under the influence of other effects. In this paper we expand on these results with a more thorough analysis of the kinematic features seen in circumbinary discs. 
We study the substructures that arise in the velocity residuals of simulated circumbinary discs. We find striking similarities between our simulations and observations of well studied transitional discs.

We structure the paper as follows: In Section~\ref{sec:method} we briefly present our hydrodynamical simulations and synthetic observations, and introduce the observational data used in this paper. In Section~\ref{sec:kin_res} we study the kinematics and residuals of simulated circumbinary discs, identifying fast flows inside the cavity and Doppler-flips as features that can appear. In Section \ref{sec:app} we then demonstrate that these features can explain two examples of kinematic residuals (vortex-like kinematics and Doppler-flips) seen in HD~142527 and HD~100546, and show fast flows are seen inside the cavities of several transition discs.
We summarise our findings in Section~\ref{sec:sum}.

\section{\textbf{Methods and Observational Data}} \label{sec:method}

\subsection{SPH Simulations and Synthetic Observations}\label{sec:rad}

We used the smoothed particle hydrodynamics (SPH; \citealt{monaghan1992}) code {\sc phantom} \citep{phantom2018}. We refer the reader to \citetalias{calcino2023} for a detailed description of the gas-only SPH models we use in the present work, however we provide a brief description here. A summary of the model parameters is provided in Table \ref{tab:ic}. Our gas discs are initialised \rr{with $5\times 10^{6}$ SPH particles following} a surface density profile $\Sigma (R)\propto R^{-p}$ for $R_\textrm{in} < R < R_\textrm{out}$, where we set $p = 1$ in all models. We assume a locally isothermal equation of state with a temperature profile $T(R) \propto R^{-2q_T}$, and $q_T=0.25$ in all models. We set the aspect ratio of the disc $H/R_\textrm{ref}$ at $R_\textrm{ref}$, with specific values listed in Table~\ref{tab:ic}. \rr{The central stars and companions are modelled as sink particles \citep{bate1995}, where for every simulation we assume a primary companion mass of 2 M$_\odot$.} The SPH artificial viscosity $\alpha_\textrm{AV}$ is used to produce a \citet{shakura1973} alpha viscosity $\alpha_\textrm{SS}$ following \citet{lodato2010}, where the value for each model is specified in Table \ref{tab:ic}. \rr{The simulations presented in this work are chosen to reflect a range of binary configurations and disc morphologies without being exhaustive. For example, circular binaries can pump large eccentricities in circumbinary discs, while eccentric binaries can produce comparatively less eccentric discs \citep{hirsh2020}. Furthermore, circumbinary discs that are co-planar, arbitrarily mis-aligned \citep{nowak2024}, or polar aligned \citep{kennedy2019}, are both theoretically and observationally supported configurations. }

Our synthetic observations are the same as shown in \citetalias{calcino2023}. We used the Monte Carlo radiative transfer code {\sc mcfost} \citep{pinte2006, pinte2009}. Since our simulations did not include dust, we assumed a power-law grain size distribution $dn/ds \propto s^{-3.5}$ for $0.03~\mu$m~$\leq~s~\leq~1$~mm with the grains well-coupled to the gas. We assumed that $T_\textrm{dust} = T_\textrm{gas}$ and all molecules are at local thermodynamical equilibrium. We only show synthetic $^{12}$CO~(3-2) line observations in the work and adopted a CO ratio of $^{12}$CO/H$_{2}=~10^{-4}$. The CO abundance is affected by photodissociation and freeze out ($T=20$~K) \citep{pinte2018}. \rr{Changing the CO line used in our analysis does not significantly alter the kinematic signatures in the velocity residuals we later study.} The primary star in each simulation has an effective temperature $T_\textrm{eff} = 8000$~K and radius $R = 1.8\ \textrm{R}_\odot$, giving a blackbody luminosity $\sim 12\ \textrm{L}_\odot$. The temperature and radius of the companions are computed from their mass listed in Table \ref{tab:ic} using the stellar and planetary tracks from \cite{siess2000} and \cite{Allard2001}. A disc inclination of $i=30^{\circ}$, a disc position angle of PA$=0^{\circ}$ and a distance of 100~pc were assumed.

The CO cubes from {\sc mcfost} were then processed to mimic the finite resolution and sensitivity capabilities of ALMA using the Python package {\sc pymcfost}. The final cubes had a spectral resolution of 250 ms$^{-1}$, and were convolved with a 0.15$\arcsec$ Gaussian beam. Artificial noise was added with a flux of $F_\textrm{Noise} = 2.5$ mJy. Integrated intensity and velocity maps were made from the cubes using {\sc bettermoments} \citep{bettermoments2018} where a cut of $5\times$ the RMS noise level was applied. We fit the velocity maps using {\sc eddy} \citep{eddy}. Since the disc flaring is relatively small in our simulations, we fit Keplerian disc models that do not account for the disc flaring. \rr{The Keplerian disc models assumed a single central massive body. We did not attempt to fit more complication rotation profiles \citep[e.g.][]{ragusa2024}.}

\subsection{Observational Data}

\begin{table*}
    \centering
    \begin{tabular}{l|c|c|c|c|c|c|c|}
    \hline
        Object &  Line & ALMA Project ID & Beam ($\arcsec$) &  $\Delta v$ (km\ s$^{-1}$) & Peak SNR$^{(\textrm{a})}$ & $r_\textrm{cavity}$ [\arcsec] & Cube Source$^{(\textrm{b})}$ \\
        \hline
        HD 163296     & $^{12}$CO (2-1) & 2018.1.01055.L & 0.15$\times$0.15 & 0.2 &  115 & - & P/PC  \\
        MWC 480       & $^{12}$CO (2-1) & 2018.1.01055.L & 0.15$\times$0.15 & 0.2 &  70  & - & P/PC \\
        GM Aur        & $^{12}$CO (2-1) & 2018.1.01055.L & 0.15$\times$0.15 & 0.2 &  34  & 0.23 & P/PC \\
        IM Lupi       & $^{12}$CO (2-1) & 2018.1.01055.L & 0.15$\times$0.15 & 0.2 &  42  & - & P/PC \\
        AS~209        & $^{12}$CO (2-1) & 2018.1.01055.L & 0.15$\times$0.15 & 0.2 &  69  & - & P/PC \\ 
        HD 97048      & $^{13}$CO (3-2) & 2016.1.00826.S & 0.11$\times$0.07 & 0.12 & 35  & 0.24 & P/PC \\
        HD~169142     & $^{12}$CO (2-1) & 2015.1.00490.S & 0.073$\times$0.067 & 0.167 & 36 & 0.23 & P/PC \\ 
                      &   &   2016.1.00344.S  &  &  &  &  & \\ 
        TW~Hya        & $^{12}$CO (2-1) &  2013.1.00387.S & 0.19$\times$0.17 & 0.04 & 275 & - & P/PC \\ 
                      &                 &  2018.A.00021.S &  &   &   &  &  \\ 
        GG Tau        &  $^{12}$CO (2-1) &  2018.1.00532.S  & 0.34$\times$0.27 & 0.08 & 88 & 1.6 & A \\
        HD 142527     & $^{13}$CO (2-1) &  2015.1.01353.S & 0.35$\times$0.31 & 0.084 & 39 &  1.18 & P/PC \\
        HD 100546     & $^{12}$CO (2-1) & 2016.1.00344.S & 0.08$\times$0.06  & 0.5  & 29 &  0.33 & P/PC \\
        CQ Tau        & $^{12}$CO (2-1) & 2013.1.00498.S & 0.12$\times$0.10  & 0.5 & 30 & 0.36 & P/PC \\
                      &  & 2016.A.00026.S &   &  &    &    & \\
                      &  & 2017.1.01404.S &   &  &    &    & \\
        AB Aurigae    & $^{12}$CO (2-1) & 2021.1.00690.S &  0.24$\times$0.17 & 0.042 & 86 &  0.98 & P/PC \\ 
        J1604         & $^{12}$CO (2-1) & 2018.1.01255.S & 0.18$\times$0.15  & 0.1 & 84 &  0.56 & P/PC \\
        LkCa 15       & $^{12}$CO (3-2) & 2018.1.00350.S &  0.07$\times$0.05 & 0.874  & 50 &  0.43 & P/PC \\ 
        CS Cha        & $^{12}$CO (3-2) & 2017.1.00969.S &  0.1$\times$0.09 & 0.25  & 59 &  0.25 & P/PC \\ 
        \hline
    \end{tabular}
    \caption{Summary of the observations used in this work. $^{(\textrm{a})}$Peak SNR refers to the peak signal-to-noise ratio seen in the channels of the cube. $^{(\textrm{b})}$P/PC: Reimaged data cube. Public data or obtained via private communication, A: Archival data product. HD~163296, MWC~480, GM~Aur, IM~Lupi, AS~209 \citep{MAPSI2021}, HD~97048 \citep{pinte2019}, HD~169142 \citep{garg2022}, TW~Hya \citep{teague2022}, GG~Tau \citep{phuong2020}, HD~142527 \citep{garg2021}, HD~100546 \citep{law2022}, CQ~Tau \citep{wolfer2021}, AB~Aurigae (Speedie et al. in press), J1604 \citep{stadler2023}, LkCa 15 (this work), CS~Cha \citep{kurtovic2022}. }
    \label{tab:obs}
\end{table*}

We utilised ALMA Band 6 and Band 7 archival $\rm ^{12}CO$ and $\rm ^{13}CO$ line data. The majority of our data was obtained either via private communication as self-calibrated data products, or via the ALMA archive, calibrated using the {\sc CASA} pipeline for the appropriate ALMA cycle. When selecting our sources we did not aim to obtain a complete or unbiased sample of discs, as we did not attempt to constrain statistical properties of the disc population. Our sample consists of discs which have been previously studied in other works \citep[e.g.][]{francis2020, wolfer2023}. 
\rr{Our sampled is based off the transition disc sampled compiled by \cite{wolfer2023}, excluding sources where the noise levels are too high, or the beam too large, to resolve the kinematics inside the cavity. We also included the MAPS sample \citep{MAPSI2021}. In a future work we will provide a detailed study of the {\sc exoALMA} sample.}
The specific details of the observations are summarised in Table \ref{tab:obs}. We imaged/re-imaged our data using the \textit{tclean} task in CASA, with the Multiscale Clean deconvolver \citep{cornwell2008} with Briggs weighting, and applied JvM \citep{JvM1995} and primary beam correction. We generated total intensity and velocity maps using {\sc bettermoments} \citep{bettermoments2018}. The total intensity maps were used to obtain the disc radius. We used both the intensity weighted average velocity (Moment~1) and quadratic methods to produce the velocity maps in this paper. 
When generating our images we applied noise cuts of $5\times$ the RMS noise for all cubes.


\section{Kinematic Signatures in Velocity Residuals}\label{sec:kin_res}

\begin{figure*}
    \centering
    \includegraphics[width=\linewidth]{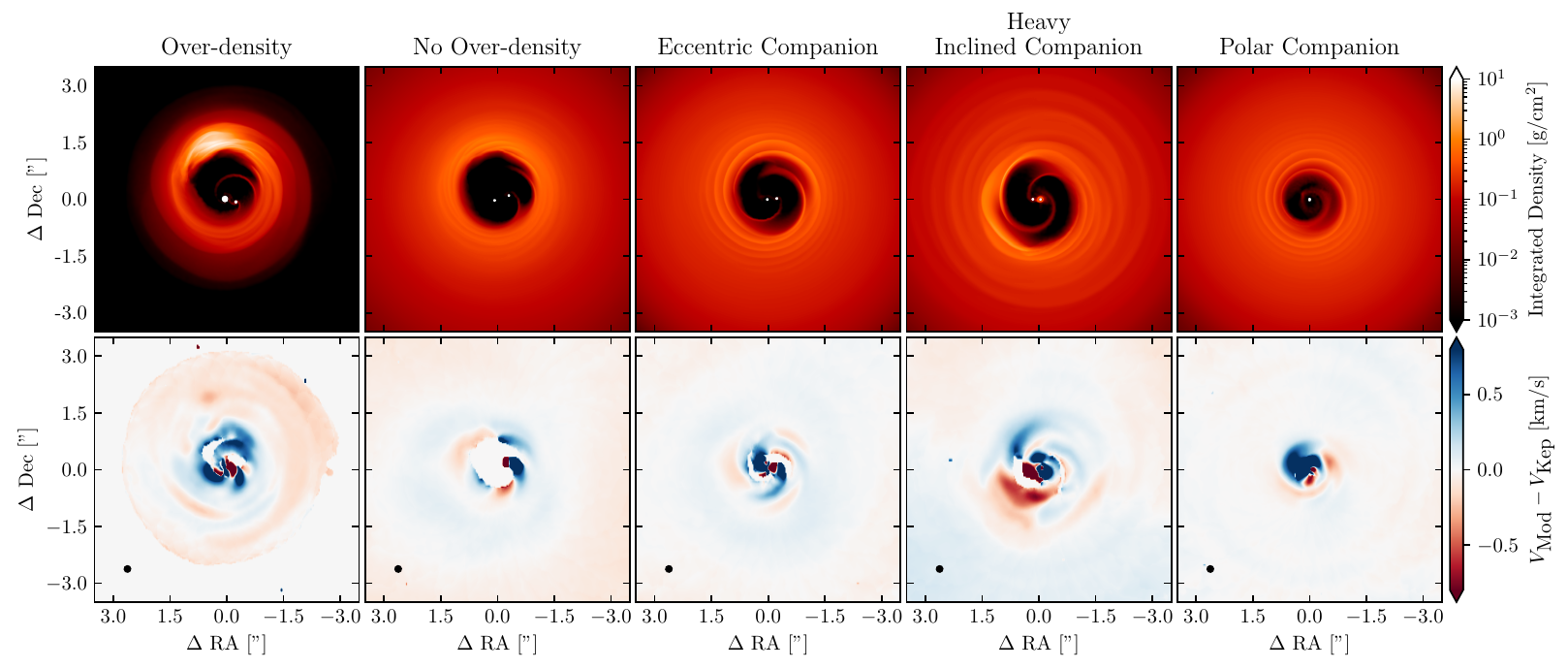}
    \caption{The surface density (top row) with the residuals from Keplerian rotation generated from CO (bottom row) for all our circumbinary disc models. Significant deviations from Keplerian rotation is observed in all models. Spiral arms and Doppler flips are common inside and co-located with the cavity. Since we used a position angle of PA$\ =0^{\circ}$ to produce our synthetic observations, the disc major axis lies along the $x$-axis. The velocity residuals have been deprojected using our fixed viewing inclination angle of $i=30^{\circ}$.}
    \label{fig:dens_res}
\end{figure*}

\begin{figure*}
    \centering
    \includegraphics[width=0.9\linewidth]{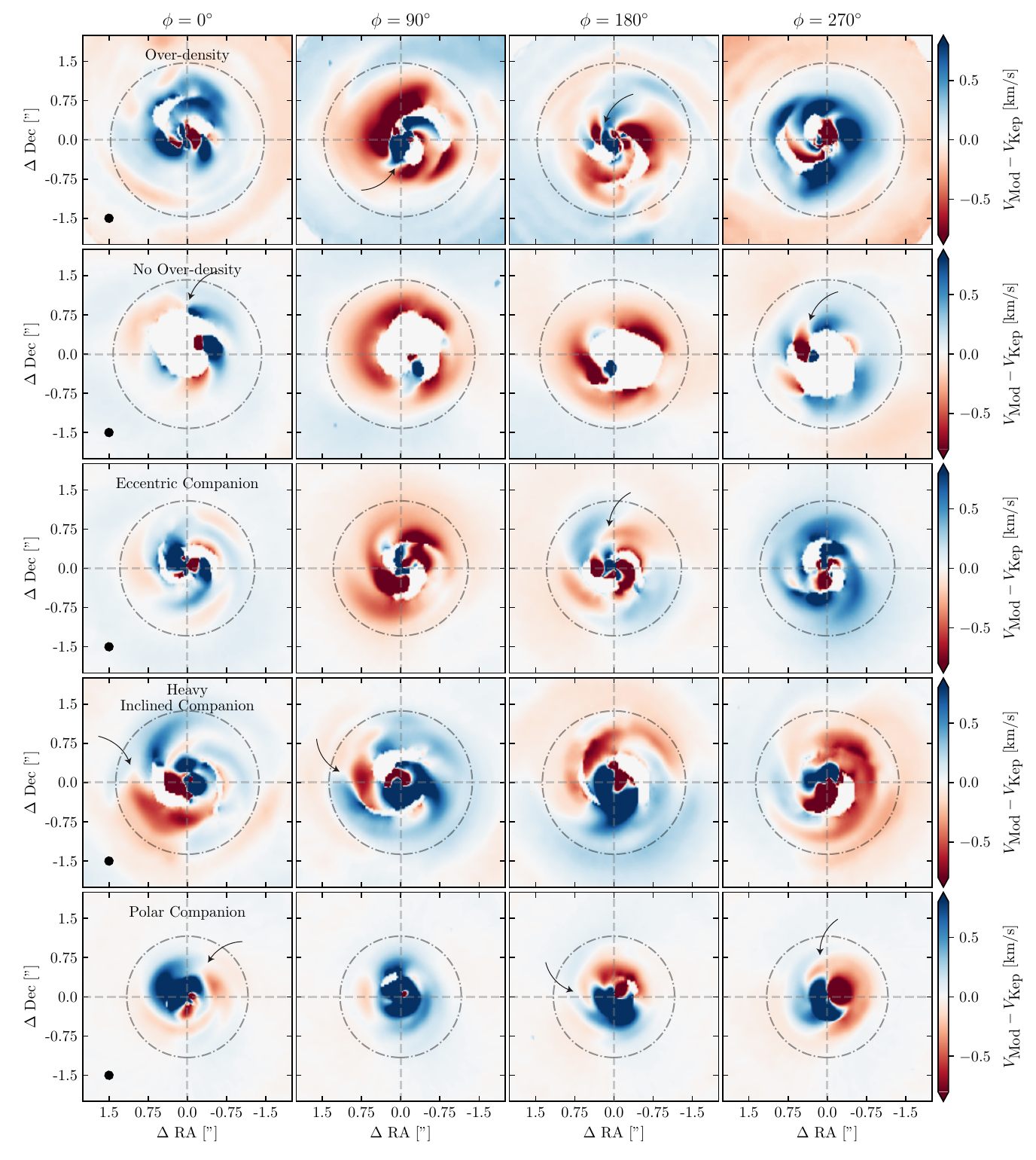}
    \caption{Binaries produce Doppler flips: Deprojected residuals from Keplerian rotation obtained using {\sc eddy} for all circumbinary models listed in Table \ref{tab:ic} for four different viewing angles (left to right columns). The horizontal and vertical grid lines are along the major and minor axis of the disc, respectively. The dash-dot circle shows the approximate radius of the cavity, obtained from the peak in the azimuthally averaged gas surface density. Doppler flips and spiral arms are ubiquitous in and around the cavity. Several prominent Doppler flips are marked with arrows.}
    \label{fig:all_res}
\end{figure*}

\begin{figure*}
    \centering
    \includegraphics[width=0.9\linewidth]{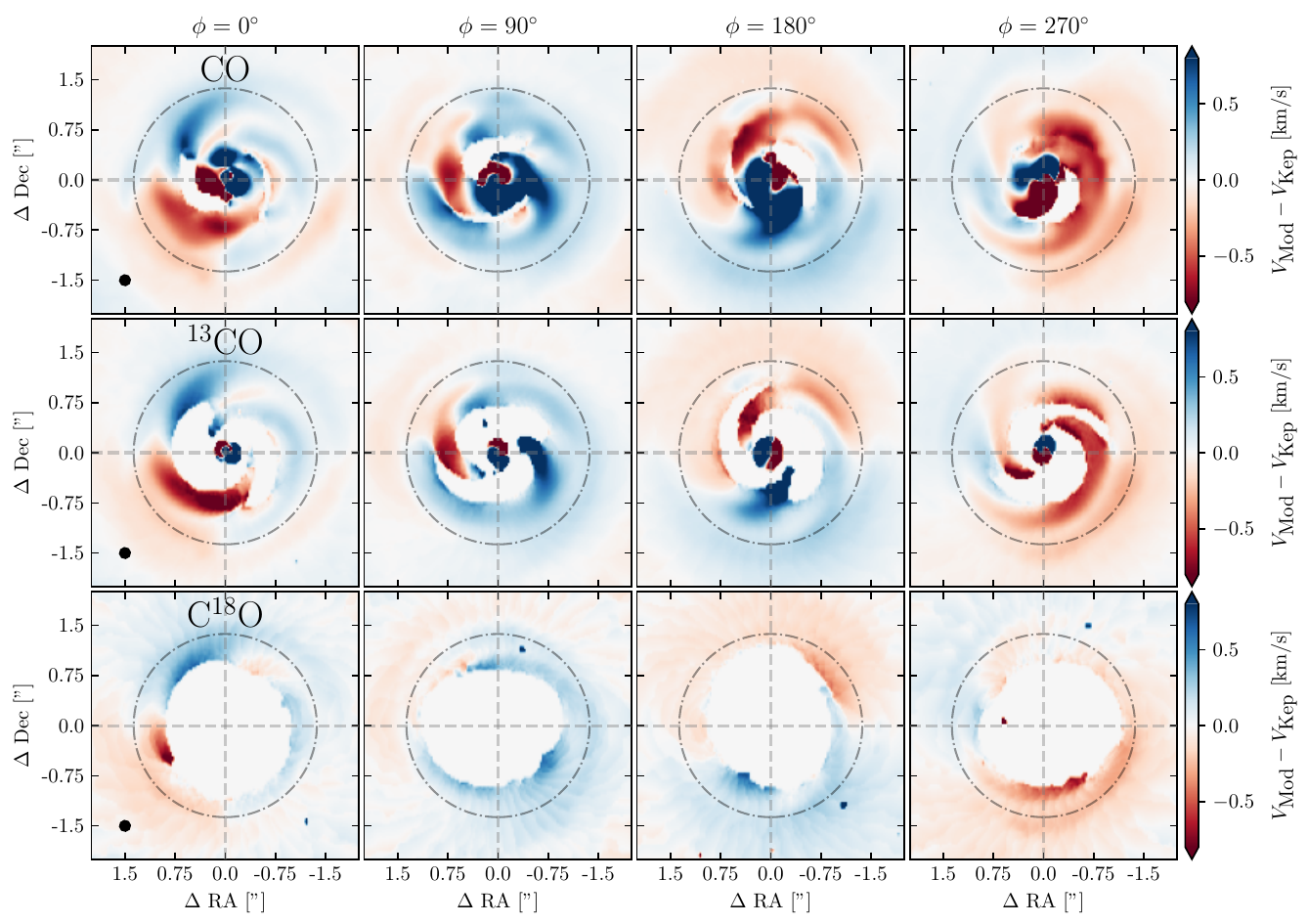}
    \caption{Deprojected residuals from Keplerian rotation obtained from the velocity map constructed from $^{12}$CO (3-2) (top row), $^{13}$CO (3-2) (middle row), and C$^{18}$O (3-2) lines. Each column is the same simulation but seen from a different viewing angle. }
    \label{fig:iso_res}
\end{figure*}

\begin{figure*}
    \centering
    \includegraphics[width=0.8\linewidth]{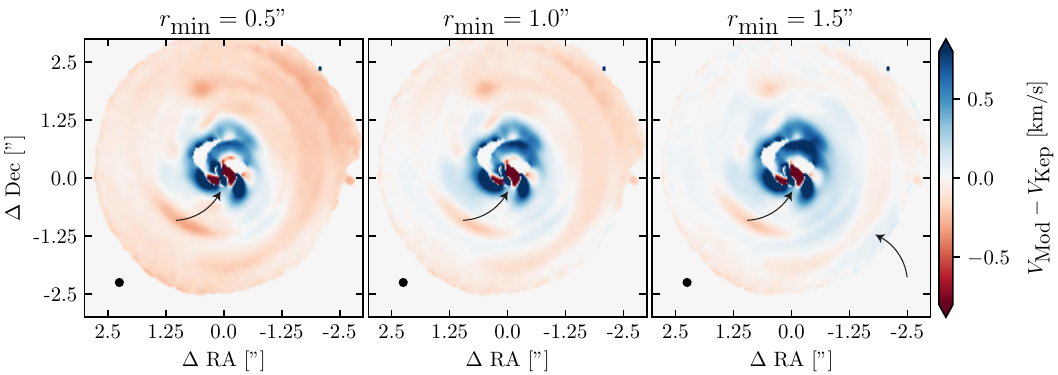}
    \caption{The deprojected residuals from Keplerian rotation for model OD with $\phi = 0^{\circ}$ with three different values of the inner radius for the fitting with {\sc eddy}. The residuals can change sign depending on the background model that is subtracted. The large amplitude residuals in the inner regions are not sensitive to the model subtracted, while the lower amplitude residuals in the outer disc are.}
    \label{fig:res_rin}
\end{figure*}

\subsection{Doppler-Flips}
We define a Doppler-flip as a change in sign of velocity across a narrow azimuthal section of the disk \citep{pp7kinematics}. 
Planet-induced Doppler-flips are produced due to a change in sign of the non-Keplerian gas motion and are centred on the planet \citep[e.g. see][]{perez2018, bollati2021, calcino2022}. Both azimuthal and radial velocities deviations are negative in front of the planet and positive behind it. Since there are both azimuthal and radial velocity perturbations producing the sign-flip, a planet at any azimuthal position in the disc can produce an observable Doppler-flip. However, since the radial velocity perturbation dominates \citep{rafikov2002, calcino2022}, the Doppler-flip should be stronger along the disc minor axis than along the major axis.

Figure \ref{fig:dens_res} shows the integrated gas surface density from the circumbinary disc simulations and the velocity residuals from their velocity maps, obtained from $^{12}$CO (3-2) line emission using the Moment~1 method. 
We identify similar signatures to planet induced Doppler-flips in the velocity residuals of our models.
In contrast to the planet induced Doppler-flips, those seen in circumbinary discs are not centred on the companion and can be a significant distance from it. Supersonic radial and azimuthal velocity perturbations are produced as the companion launches spiral arms and interacts with gas trying to flow inside and fill the cavity. They can also have sudden changes in sign as both the primary and secondary stars launch spiral arms into slower moving material. These large perturbations can persist outside of the gas depleted central cavity and be spatially co-located with the peak in the background gas surface density, which is where we expect large dust grains to accumulate. 

Figure \ref{fig:all_res} shows the velocity residuals for our circumbinary disc models at different viewing angles (labelled $\phi$). Sudden changes in the sign of the velocity residuals --- Doppler-flips --- are abundant inside and around the cavity. Arrows in Figure \ref{fig:all_res} indicate several prominent Doppler-flips. \rr{We also mark the approximate radial location of the gas pressure maximum around the cavity with a dot-dash circle. The location of the Doppler-flips} relative to the cavity also suggests they could be co-located with other features attributed to circumbinary discs, such as dust rings or dust asymmetries. 
Whether or not a Doppler-flip is seen can depend on the disc model that is subtracted. In Figure \ref{fig:res_rin} we show the velocity residuals of model OD with three different Keplerian models which were obtained by changing the minimum radius, $r_\textrm{min}$, that is used in the fit with {\sc eddy}. \rr{The disc model in {\sc eddy} assumes the gas is on a circular Keplerian orbit around a single star}. Although the high amplitude Doppler-flips (shown with an arrow in all three columns) do not change sign, lower amplitude residuals, particularly in the outer disc (bottom right of right-most panel), can change sign. \rr{In principle this is not just an issue when fitting the velocity map of known circumbinary discs, but also for single stars \citep[e.g.][]{teague2022}. By their very definition velocity residuals are model dependent, and the appearance of a Doppler-flip is therefore also model dependent. However high amplitude Doppler-flips are less dependent on the Keplerian model that is subtracted, as shown in Figure \ref{fig:res_rin}. } 

\rr{These results demonstrate that the appearance of Doppler-flips, particularly low velocity (i.e. order few hundred meters per second) are sensitive to the disc model that is subtracted. Since we assume a Keplerian background model, large residuals should be expected. However what is notable is that the large residuals can still be visible in proximity to the cavity edge and gas pressure maxima. For example, in the HIC model (fourth row, first column of Figure \ref{fig:all_res}) a Doppler-flip on the order of several hundred meters per second is seen just interior to the gas pressure maxima at a radial location exterior to the binary orbit.}

\rr{The Doppler-flips we observe in our simulations are also sensitive to the CO isotopologue they are observed with, as shown in Figure \ref{fig:iso_res}. In Figure 9 of \citetalias{calcino2023}, using the same simulation we use here, we showed how velocity kinks seen in the channel maps also depend on the CO isotopologue. In that Figure, the apparent amplitude of the kink decreases due to the more highly perturbed material responsible for the kink becomes unobservable due to the lower abundance. The same effect results in the Doppler-flips becoming less apparent in the residual maps of Figure \ref{fig:iso_res}.}

The ubiquity of Doppler-flip features in a circumbinary disc suggest that they should not be used as an indicator of planet-disc interactions, particularly when they are \rr{in proximity of} a gas and dust depleted cavity in a transition disc. \rr{Contrary to those generated by planets, Doppler-flips in CBDs are not necessarily directly associated with a companion, but rather they are a result of spiral structure exterior to the binary orbit.} The structure of the Doppler-flips we observe are reminiscent of the one in HD~100546 reported by \cite{casassus2019}, which we comment more on in Section \ref{sec:hd100}.

\subsection{Spiral Arms: The Origin of Doppler-Flips}

\begin{figure*}
    \centering
    \includegraphics[width=0.6\linewidth]{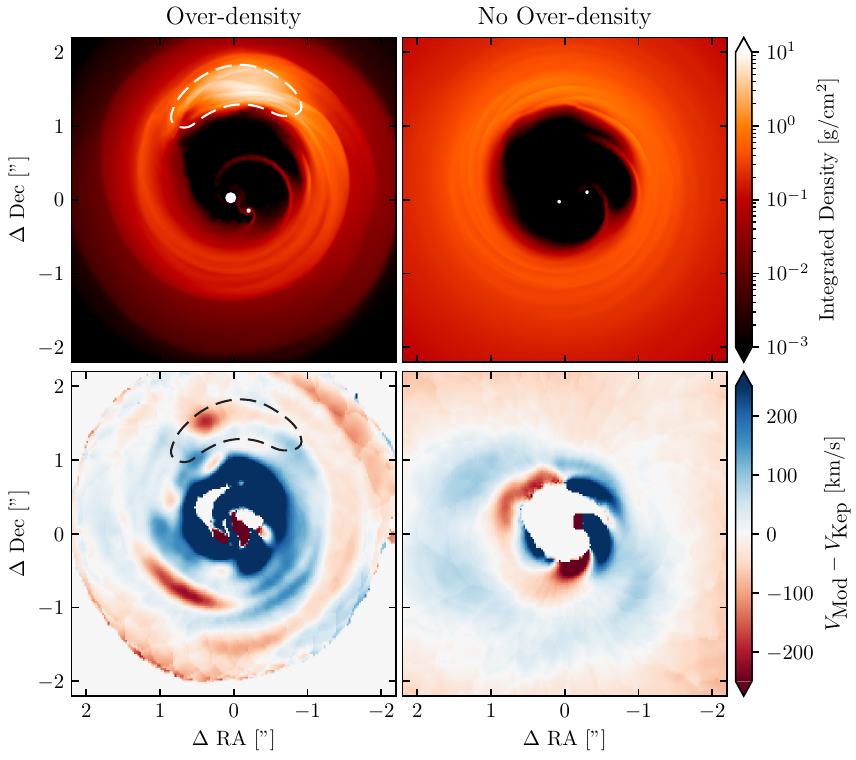}
    \caption{The surface density (top row) with deprojected residuals from Keplerian rotation generated from CO (bottom row) of models over-density (OD) and no over-density (NOD). The over-density produces a Doppler-flip signature across it's major axis. There is also additional spiral structure through the disc with OD than in no OD model, which has a very similar cavity size and shape as model OD. These additional spiral are arising due to the over-density feature, and are apparent in the velocity residuals.}
    \label{fig:dens_res_com}
\end{figure*}

Circumbinary discs are rich with spiral structure. These spirals are visible in the velocity residuals. 
Planet-induced Lindblad spirals appear mostly as radial perturbations and the radial velocity changes sign across the spiral \citep{bollati2021, calcino2022}.
The spiral arms induced by stellar mass companions produce larger perturbations in the velocity field of the gas than their planetary counterparts, although they are still mostly generated by Lindblad resonances. 
The spiral structure is most abundant near and inside the cavity. Since this area is also typically associated with (sub-)mm dust continuum emission, \rr{spiral structures may also be imprinted on the dust continuum emission. Planet induced spiral wakes are visible in the mm dust continuum emission if the dust and gas are well coupled \citep[e.g.][]{speedie2022, verrios2022}. \citet{poblete2019} showed that marginally decoupled dust grains can produce dusty clumps around the central cavity. It is plausible that the binary induced spiral structure may be imprinted in the mm dust continuum if the grains are sufficiently coupled to the gas.}

\rr{In \citetalias{calcino2023} we showed that spiral structure in and around the central cavity of a CBD produces substantial velocity perturbations away from Keplerian rotation. These perturbations are manifested as large residuals after subtracting the best fit Keplerian disc model, and occasionally produce the Doppler-flips we discussed in the previous section. Given the causal link between spiral arms and Doppler-flips, we should then expect that a Doppler-flip observed on the edge of a cavity, produced by an inner binary, should be accompanied by spiral structure. }

Spiral structure not only originates directly from the binary interacting with the disc through Lindblad resonances, but also from the over-density that can develop. 
In the left most column of Figure \ref{fig:dens_res_com} we show our model with an over-density (model OD) in integrated surface density with the corresponding velocity residuals below. The over-density is highlighted by white and black dashed lines in the surface density and velocity residuals, respectively. There is additional spiral structure through the disc that is not seen in the no over-density (NOD) model, shown in the right column, which has a very similar cavity size and shape as model OD. 
\textbf{The formation} of the over-density is likely related to the growth of the eccentric cavity, where a disc initialised with a smaller cavity leads to a more prominent over-density developing \citep[see Section 4.3 of][for more details]{ragusa2020}. \rr{Thus the spirals are a direct result of the over-density, and not due to the eccentric cavity.}
These spirals propagate far from the cavity, and are also apparent in the velocity residuals. \rr{Since our model does not include gas self-gravity, the spiral structure produced from the over-density is not due to the gravitational potential of the over-density exciting spiral wakes \citep[e.g. see][]{zhu2016}. We note that a similar trailing spiral is seen in the CBD simulations of \cite{rabago2023}. The spiral from the over-density is generated by hydrodynamical effects: the over-density is a localised region of higher density and pressure which perturbs the surrounding disc material as it orbits, resulting in the additional spiral structures.} We show in Section \ref{sec:hd14} that the over-density might be responsible for the extended spiral structure seen in HD~142527.

\subsection{Eccentric Cavity}

In three of our models the gas orbiting around the cavity has a non-negligible eccentricity which appears in the velocity residuals. 
These are models over-density (OD), no over-density (NOD), and eccentric companion (EC). 
Large red and blueshifted residuals become more apparent when the eccentricity vector of the disk is pointing perpendicular to the line of sight, which is the case when $\phi = 90^{\circ}$ 
and $\phi = 270 ^{\circ}$ in Figure \ref{fig:all_res}. 

The residuals from eccentric discs arise due to the sub and super-Keplerian motion that occur at apoapsis and periapsis, respectively. When fitting a velocity map is it common to mask the central regions \citep[e.g.][]{teague2021, teague2022}. For a disc with a cavity and substantial deviations from Keplerian rotation, these non-Keplerian motions are also often masked out \citep[e.g.][]{boehler2021, wolfer2021, norfolk2022}. As we show in Figure~\ref{fig:res_rin}, the final residual map that is obtained depends on the model that is subtracted, which in our eccentric disc models is sensitive to the minimum radius assumed. This issue could be alleviated by fitting an eccentric disc rotation profile to the observations \citep[e.g.][]{kuo2022, yang2023}. For a circumbinary discs, we expect that the disc eccentricity decreases as a function of disc radius, and hence only the regions close to the cavity may be highly eccentric. 

Aside from the kinematics \citep{kuo2022}, an eccentric cavity might also be inferred from the spatially resolved distribution of gas 
(through line emission observations) and dust (through scattered light or thermal emission observations; \citealt{dong2018, yang2023}). 
If the location of the primary star can be determined, deprojecting the spatially resolved observations can 
also lead to constraints on the disc eccentricity \citep{yang2023}.

\subsection{Fast Flows Inside the Cavity}

\begin{figure*}
    \centering
    \includegraphics[width=1.0\linewidth]{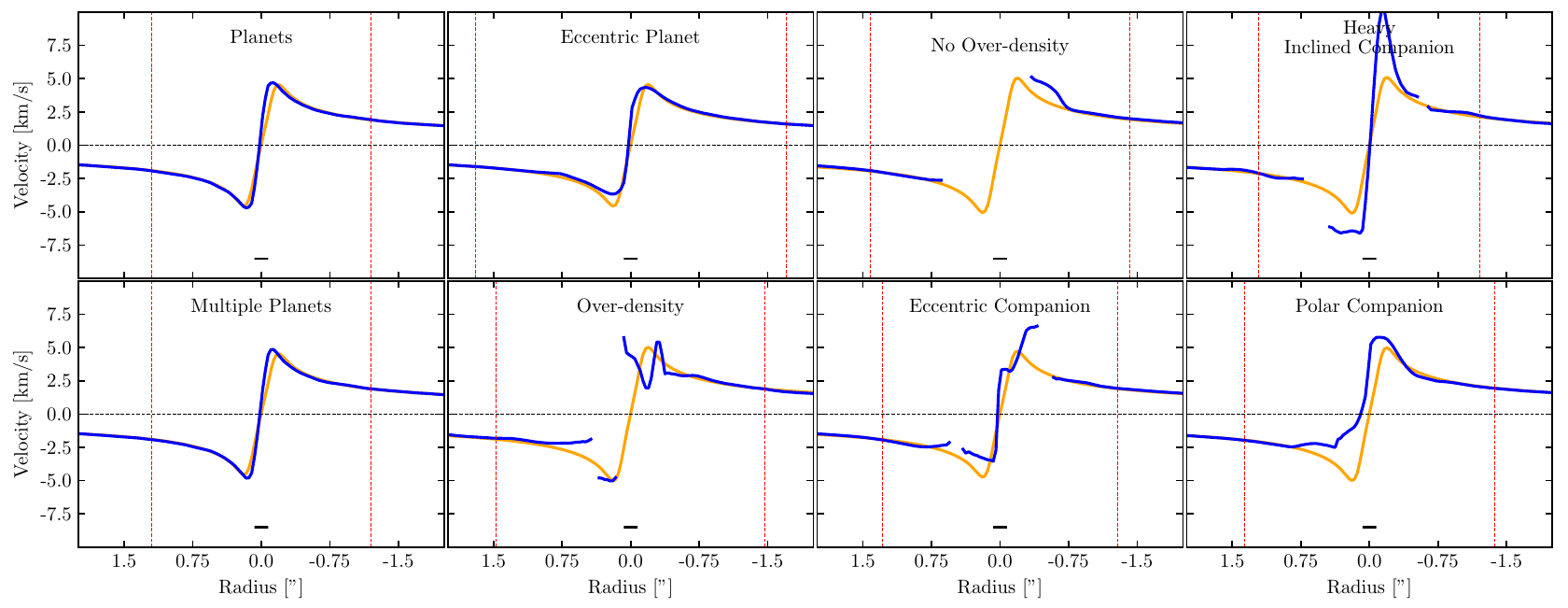}
    \caption{The rotation curve along the disc major axis of each model presented in Table \ref{tab:ic} (blue lines) with the expected Keplerian rotation of the unperturbed disc (orange curve).
    The vertical red lines mark the location of the edge of the cavity as seen with C$^{18}$O (Figure 5 of Paper I). Minor deviations from Keplerian rotation are seen in the Planet and Multiple Planets models, mostly due to the assumptions used to generated the Keplerian rotation curve.
    Modest deviations from Keplerian rotation are seen inside the cavity of the Eccentric Planet model, while in most of the circumbinary disc models several km/s deviations are seen. Since radial velocities are largely perpendicular to the line of sight along the major axis, the deviations from Keplerian rotation are mostly due to azimuthal perturbations. The black horizontal line represents the size of the beam major axis.}
    \label{fig:rot_curve}
\end{figure*}

As we showed in \citetalias{calcino2023}, and in Figure~\ref{fig:dens_res}, both radial and azimuthal velocity perturbations increase inside the cavity of a circumbinary disc. 
For the inclined companion models, non-negligible vertical motion is also observed.
After subtracting a best-fit Keplerian rotation field, these fast flows are obvious in the residuals (Figure~\ref{fig:dens_res})
and span the entire width of the cavity. In some instances, the residuals change sign across the major 
axis of the disc (e.g. first column of model NOD in Figure~\ref{fig:all_res}), indicating mostly radial flows \citep[e.g.][]{teague2019} (e.g. model HIC), while in other instances a 
superposition of azimuthal and radial velocities lead to no obvious sign change across either the major or minor axes (e.g. multiple viewing angles of models OD and NOD).

To further explore how these fast flows appear in observations, Figure \ref{fig:rot_curve} shows the 
velocity along the major axis of the disc neglecting any disc flaring with the blue curves, while the 
orange curves show the expected mid-plane Keplerian rotation neglecting any disc flaring and gas pressure support. Along the major axis the radial velocity component 
of a disc is negligible since the major velocity component is perpendicular to the line of sight. For the planet simulations (model P and MP), the deviations from the estimated Keplerian rotation are small. In contrast, the eccentric planet model and the circumbinary disc models show larger deviations. These perturbations span several beams 
and hence are not localised. They also exceed estimates of the local sound speed by a factor of several. Highly perturbed rotations curves inside of a cavity can thus indicate that the disc is circumbinary.
Similar perturbations are seen in the rotation curve and residuals inside of the cavity of GG Tau and AB Aurigae (\citealt{Riviere2019}; see also Section \ref{sec:obsff}). 

\subsection{Vortex-like Kinematic Signatures}

\begin{figure*}
    \centering
    \includegraphics[width=\linewidth]{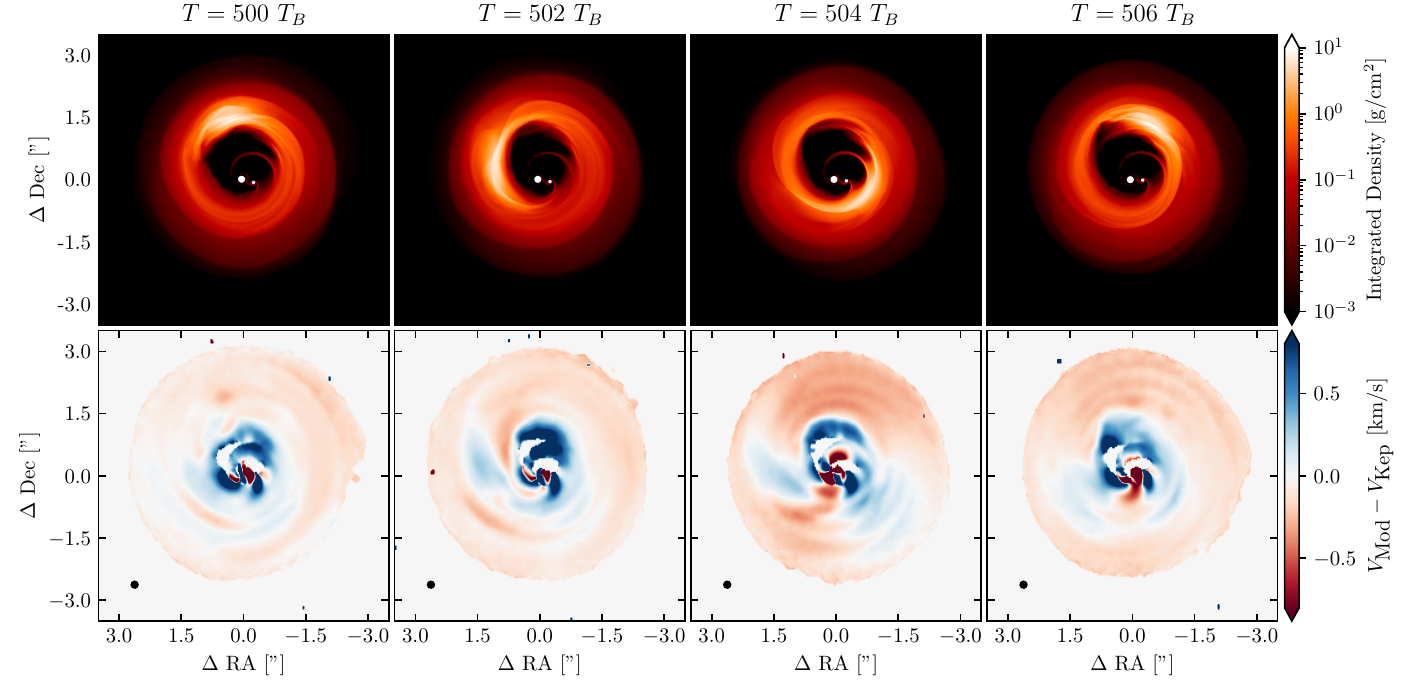}
    \caption{The surface density (top row) moment 1 (top row) with deprojected residuals from Keplerian rotation generated from CO for model OD at different time snapshots. Significant deviations from Keplerian rotation is observed in all models. Spiral arms and Doppler flips are common inside and co-located with the cavity.}
    \label{fig:dens_res_irs}
\end{figure*}

Figure \ref{fig:dens_res_irs} shows the surface density and velocity residuals of the over-density model for four different snapshots which trace the orbital phase of the over-density. \rr{The over-dense feature arises as the eccentric cavity grows around the binary \citep{ragusa2020}, and is distinct from a vortex as it lacks vortical motion \citep{ragusa2017}. Since it orbits around the cavity at the Keplerian frequency, it is also not a ``traffic-jam'' of dust grains at the apocenter of an eccentric disc \citep{ataiee13a}}. In \citetalias{calcino2023} we showed that the over-dense feature orbiting a circumbinary disc also shows a change in velocity 
across the major axis of the over-density. Since the residuals are dominated 
by excess velocities arising from the eccentric gas motion around the cavity, the signature from the over-dense 
feature can be difficult to distinguish.

In left column of Figure \ref{fig:dens_res_com} we see that this change 
in velocity is apparent in the velocity residual. The over-density is outlined with a dashed line in the integrated surface density and velocity residual plots. A change in sign across the major axis of the over-density is apparent. The inner binary clearly complicates any analysis of the kinematics. This fact demonstrates that even with high spectral and spatial resolution 
observations, disentangling the velocity signature of the over-dense feature, or `vortex' as in \cite{boehler2021}, is difficult. 

\section{Applications to Observed Kinematic Features} \label{sec:app}

\subsection{HD~142527}\label{sec:hd14}

\begin{figure*}
    \centering
    \includegraphics[width=0.9\linewidth]{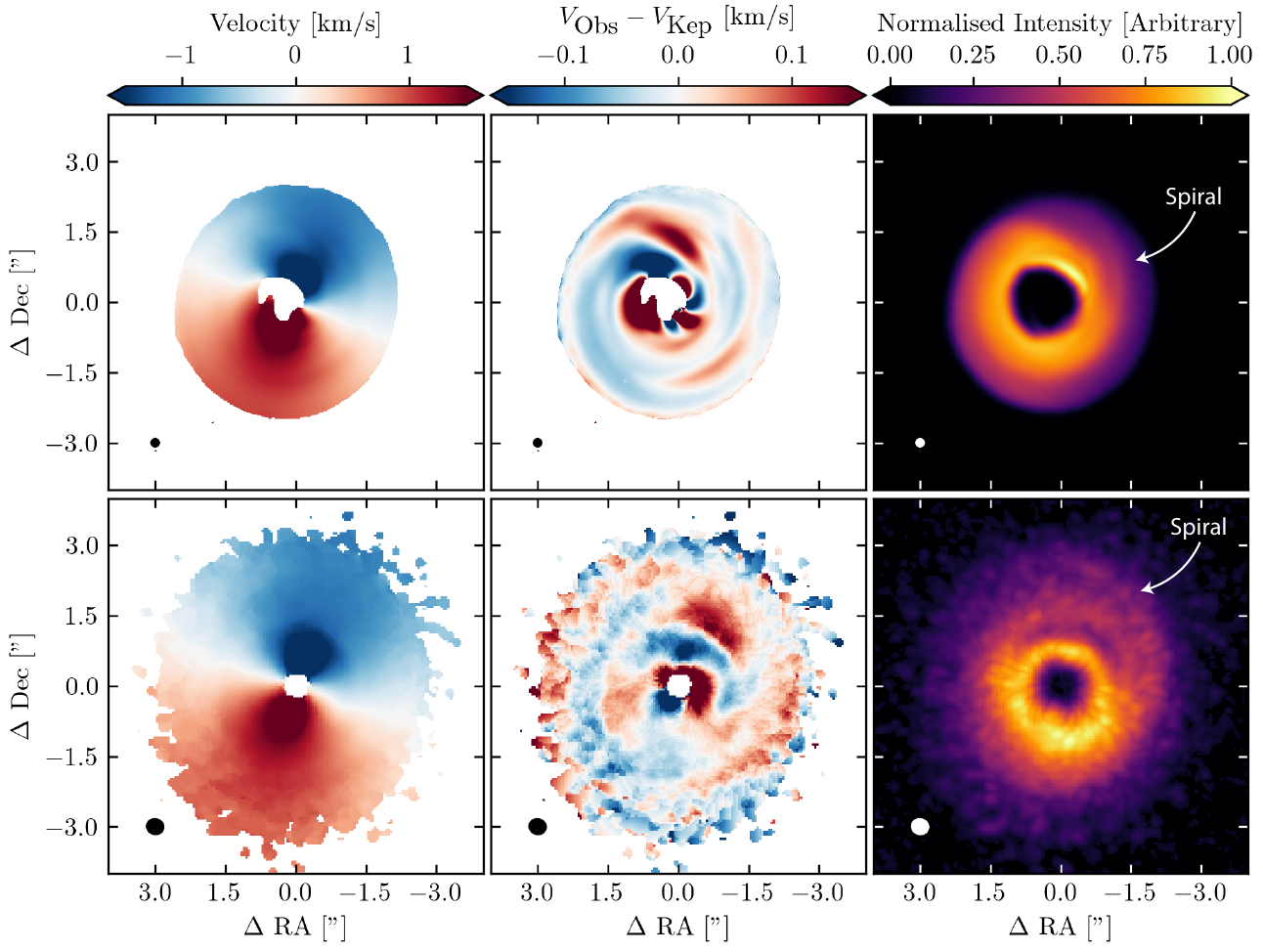}
    \caption{The moment 1 (left column), moment 1 residuals from a best-fit Keplerian disc model (middle column), and the peak intensity (right column) of our over-density model (top row) and HD~142527 (bottom row). The over-dense feature can reproduce the spiral structure seen in HD~142527 in both peak intensity and in the velocity residuals. The location of the over-dense feature is similar in both the model and observations. A Doppler flip across the over-density is reproduced in our model.}
    \label{fig:hd14}
\end{figure*}

HD~142527 is a well studied binary system containing a 0.1-0.4~M$_\odot$ companion \citep{biller2012, close2014, lacour2016, christiaens2018}. It also displays non-Keplerian gas flows 
\citep{casassus2015, garg2021} inside of its $\sim$140~au cavity \citep{casassus2013, avenhaus2017}. 

This binary was 
subject to the study by \cite{price2018}, where it was shown that many of the morphological features in this disc can 
be explained by the observed companion, assuming a rather large binary semi-major axis of $\sim 30-40$ au. Recently, \cite{nowak2024} constrained the semi-major axis of HD~142527B to $\approx 10$ au, which is too small to completely explain the observed cavity size. 
Nevertheless, spiral arms have been observed in the outer disk in CO integrated emission and peak intensity 
\citep{christiaens2014, garg2021}. They are also seen after subtraction of a best-fit Keplerian disk model by \cite{boehler2021}, 
along with kinematic signatures hypothesised to originate from anti-cyclonic motion (which we refer to as ``vortical motion'') inside the dust trap \citep{casassus2013, ohashi2008}. 
We have shown that such kinematic features can also be seen in a circumbinary over-density, and may explain both this feature and the spiral arms.

To demonstrate this, we rescale our model OD such that the primary star has a mass of 1.8 M$_\odot$, consistent with the mass of 
HD~142527 \citep{fukagawa2006}. The length is rescaled by a factor of 3.8 from the original model presented in \cite{calcino2019} 
such that the total diameter of the cavity along the major axis is $\sim 280$ au. When performing our radiative transfer 
calculations we closely follow the prescription summarised in Section \ref{sec:rad} with a few differences. We set the inclination 
to $i=27^{\circ}$ and position angle $\textrm{PA}= -20 ^{\circ}$ \citep{avenhaus2014, perez2015b}. The temperature and radius of the 
central star are set to 6500 K and 2.8 R$_\odot$ \citep{Mendigutia2014}. For the companion we assume a temperature of $3000$ K and 
radius of 0.9 R$_\odot$ \citep{lacour2016}. We also scale the total gas mass of our simulation to be consistent with the gas surface 
density estimates from \cite{garg2021}. 

\begin{figure*}
    \centering
    \includegraphics[width=0.7\linewidth]{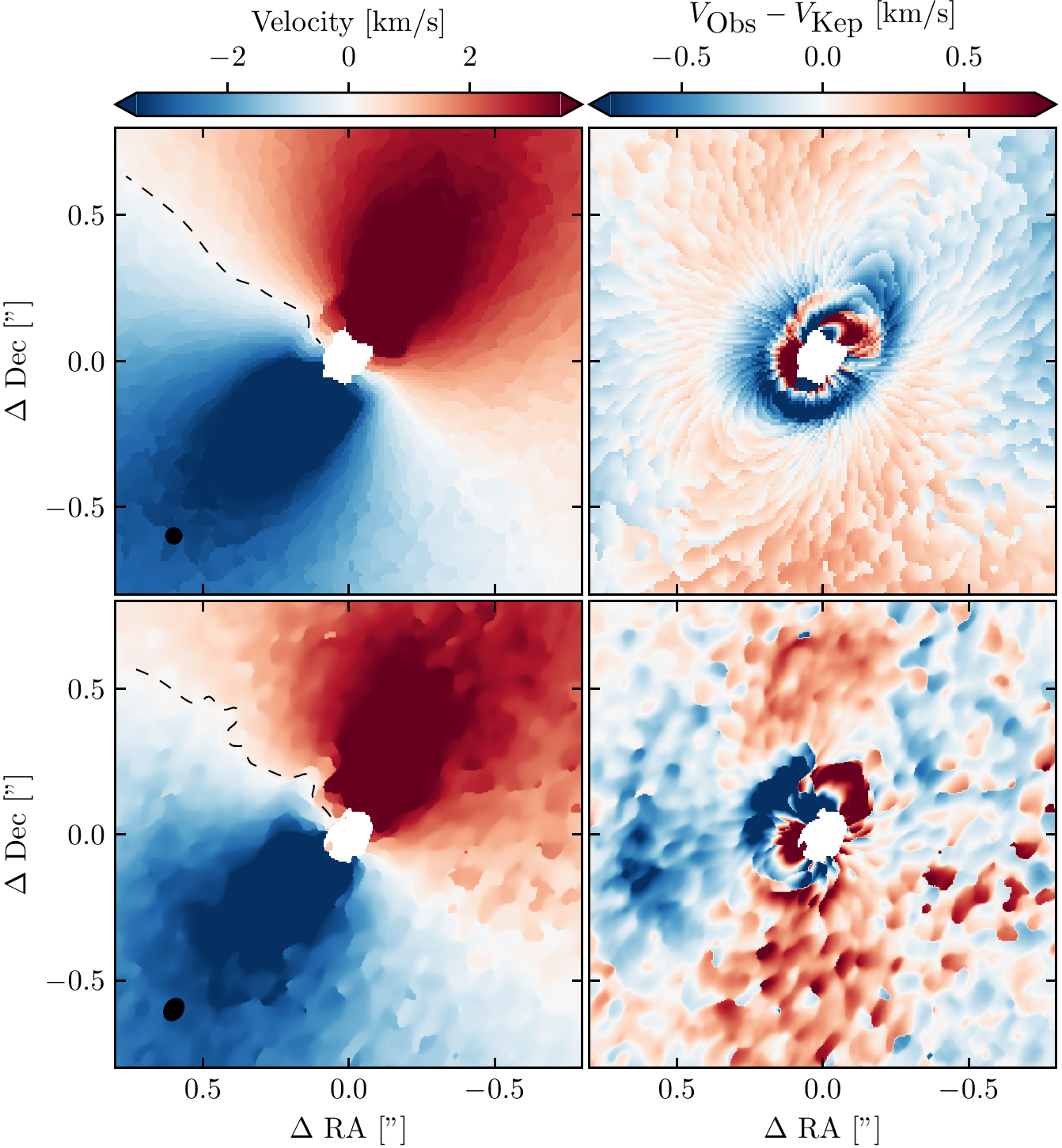}
    \caption{The velocity map produced with the quadratic method (left column) and the residuals from a best fit rotation profile (right column) of our eccentric disc model (top row) and HD~100546 (bottom row). Our eccentric disc model reproduces the kink in the velocity map and the excess in velocity residuals around the cavity of the observations. We produce the Doppler flip signature with our eccentric companion model.}
    \label{fig:com_hd10}
\end{figure*}

We obtained the ALMA $^{13}$CO data (2015.1.01353.S, PI Christiaens) used in \cite{garg2021} and \cite{boehler2021}. We show the velocity map, 
velocity map residuals, and $^{13}$CO peak intensity of this data and our model in Figure \ref{fig:hd14}. Firstly, we see that our model recreates the 
observed spiral arms in the outer region of the disc found by \citep{christiaens2014} and further studied by \citep{garg2021}. These spirals are 
associated with perturbations in the velocity field of the disc (see Figures 1 and 4 in \citetalias{calcino2023}) and appear in our velocity residuals. These are 
also seen in the data of HD~142527, as reported by \cite{boehler2021}. Our model is also able to recreate the Doppler-flip co-located with the 
over-density. \cite{boehler2021} suggest this feature is arising due to vortical motion inside the dust trap in HD~142527, however our 
results show that such a feature can be create without any vortical motion. Indeed, while analysing our model we found that Doppler-flips close 
to the cavity edge are exceptionally common, with some co-located with the over-density while others are not. This should not be too 
surprising given the complex kinematics seen in our simulations in Figures 1 and 3 of \citetalias{calcino2023}.

Recently, \cite{rabago2023} studied misaligned circumbinary discs using the grid code {\sc Athena++}, allowing them to have fine control over the disc viscosity. 
Most previous studies employ smoothed particle hydrodynamics \citep[e.g. ][]{price2018}, where it is more difficult to study circumbinary discs with low ($\alpha < 10^{-4}$) viscosity. 
The results from \cite{rabago2023} mostly agree with previous SPH results regarding disc breaking, warping, and cavity size. 
Interestingly in their $\alpha = 10^{-5}$ simulation a long-lived vortex is formed in the circumbinary disc. 
The anti-cyclonic motion inside the vortex is distinct from the mostly solid-body motion of the lump seen in SPH simulations. 
However it is unclear whether it would be possible to distinguish between these two scenarios using disc kinematics alone as the velocity residuals of a circumbinary disc are
already complicated. We note that the simulations by \cite{rabago2023} also cannot explain the observed cavity size given the $\sim 10$ au semi-major axis of HD~142527B. A low disc viscosity alone cannot explain the large cavity size. Another companion inside the cavity, or including additional physics in the simulation, may resolve this issue. 

We summarise that the spiral arms in CO integrated emission and peak intensity, along with their associated kinematic signatures, 
may directly arise from the horseshoe feature in HD~142527. It is unclear whether the kinematics of the horseshoe feature can be interpreted as a vortex in a circumbinary disc, as they 
are also consistent with an over-dense feature lacking vortical motion. In light of the recent results of \cite{nowak2024}, it is not clear if there is a direct causal link between the dust over-density and the inner binary in HD~142527.

\subsection{HD~100546}\label{sec:hd100}

The Herbig Ae/Be star HD~100546 is surrounded by a transition disc which displays all of the morphological features we can attribute to a circumbinary disc. 
Firstly, it displays a cavity of radius $\sim 20$ au in the $\sim$mm dust continuum which extends out to $\sim 40$ au \citep{wright2015, pineda2019, perez2020, norfolk2021}, \rr{with a faint ring located at 200 au \citep{walsh2014, fedele2021}}. 
The \rr{most central dust} continuum \textbf{ring} also shows an asymmetry, which in the higher resolution observations of \cite{perez2020} resemble tightly wound spiral arms. 
A cavity is also present in the $^{12}$CO, implying it is heavily depleted in not just large dust grains, but also gas \citep{pineda2019, perez2020}. Spiral arms are also observed in 
scattered light close to the cavity and \textbf{on larger scales} at the edge of the protoplanetary disc \citep{sissa2018, fedele2021}. Finally, velocity kinks are seen in the $^{12}$CO 
iso-velocity curves outside of and on the edge of the cavity \citep{perez2020}, along with a Doppler-flip on top of the $\sim$mm continuum ring \citep{casassus2019}. 
Kinematic perturbations are also observed inside the cavity around the $\sim$mm continuum emission associated with the central star \citep{perez2020}. \cite{casassus2019} and \cite{perez2020} attributed these kinematic signatures to a 5-10 M$_\textrm{J}$ planet, and 
suggested the Doppler-flip on top of the continuum emission results from gas flows close to the planet. This scenario is in tension with the position 
of the continuum emission; such a large planet will carve a gap and produce a pressure maximum beyond its orbital location. Dust grains will accumulate 
inside this pressure maximum and thus we should reasonably expect a Doppler-flip signature to be seen well inside the dust cavity, and not on top of the 
dust continuum emission.

Figure 1 of \cite{casassus2019} shows the deprojected continuum emission with the Doppler-flip superimposed. From this perspective, the 
Doppler-flip looks to be associated with the spiral structures in the continuum ring. Several of our circumbinary disc models do show sharp 
changes in gas velocity associated with spiral arms on the cavity edge, potentially explaining this feature in HD~100546. To demonstrate this 
we obtained the CO J=2–1 from \cite{perez2020}, made publicly available from \cite{law2022}.
We also \rr{reproduced} our eccentric companion model with \rr{a new simulation with} a binary semi-major of 10 au, and eccentricity of 0.4. \rr{We produced synthetic CO J=2-1 observations assuming stellar parameters similar to those in HD~100546 \cite[$T=9750$ K, M$_{\star}=2$ M$_\odot$][]{vioque2018}}.

We produce velocity maps using the 
quadratic method \citep{teague2018} and compute their residuals using {\sc eddy}. These are shown in Figure \ref{fig:com_hd10} for our model (top row) 
and the observations (bottom row). As indicated by the dashed black line, which marks the $v_\textrm{los} = 0$ km/s iso-velocity curve, the velocity kink seen in the channel maps \citep[see][]{perez2020, norfolk2022} also appears in the velocity map. 
Aside from the Doppler-flip reported by \cite{perez2020}, there also appear to be other significant velocity perturbations further inside the dust cavity. The Doppler-flip signature along with the other perturbations inside the cavity are reproduced in our eccentric companion model. 

Recently \cite{casassus2022} proposed that the Doppler-flip signal in HD~100546 could be due to a disc eruption event or inflow onto the disc. The latter hypothesis seems rather unlikely, as it is not clear how inflowing material could produce both blue and redshifted emission. The disc eruption scenario also lacks a physical mechanism to explain the eruption, and whether such a scenario could produce both blue and redshifted super-sonic perturbations that span roughly half the azimuthal extent of the disc (and appear inside of the cavity) is not demonstrated. The association of a Doppler-flip with spiral structure seen in the continuum emission \citep{casassus2022}, as well as emission from the shock tracing molecule SO \citep{booth2018, booth2023}, is expected with our scenario of a massive internal companion. 

The main criticism of our proposal is that such a companion probably should have been detected in the sparse aperture masking observations with VLT/SPHERE by \cite{perez2020}, and more recently by \cite{stolker2023}. However as noted in \cite{norfolk2022}, the observations by \cite{perez2020} conducted in May 2018, and more recently throughout 2021 and 2022 by \citep{stolker2023}, were conducted when a suspected companion detected by \cite{brittain2014} was predicted to be behind the near-side of the disc \citep{brittain2019}. Interestingly, \cite{stolker2023} find a bright spot on the near-side of the disc that is in close proximity to where one would expect this potential companion to be located. However they conclude this bright spot is due to bright forward scattering on the near-side of the disc. Additional sparse aperture masking observations tracking these bright spots may help to constrain their orbital motion and determine if either are consistent with predictions made by \cite{brittain2014, brittain2019}. 

\subsection{Fast-Flows Inside the Cavities of Transition Discs}\label{sec:obsff}

\begin{figure*}
    \centering
    \includegraphics[width=1.0\linewidth]{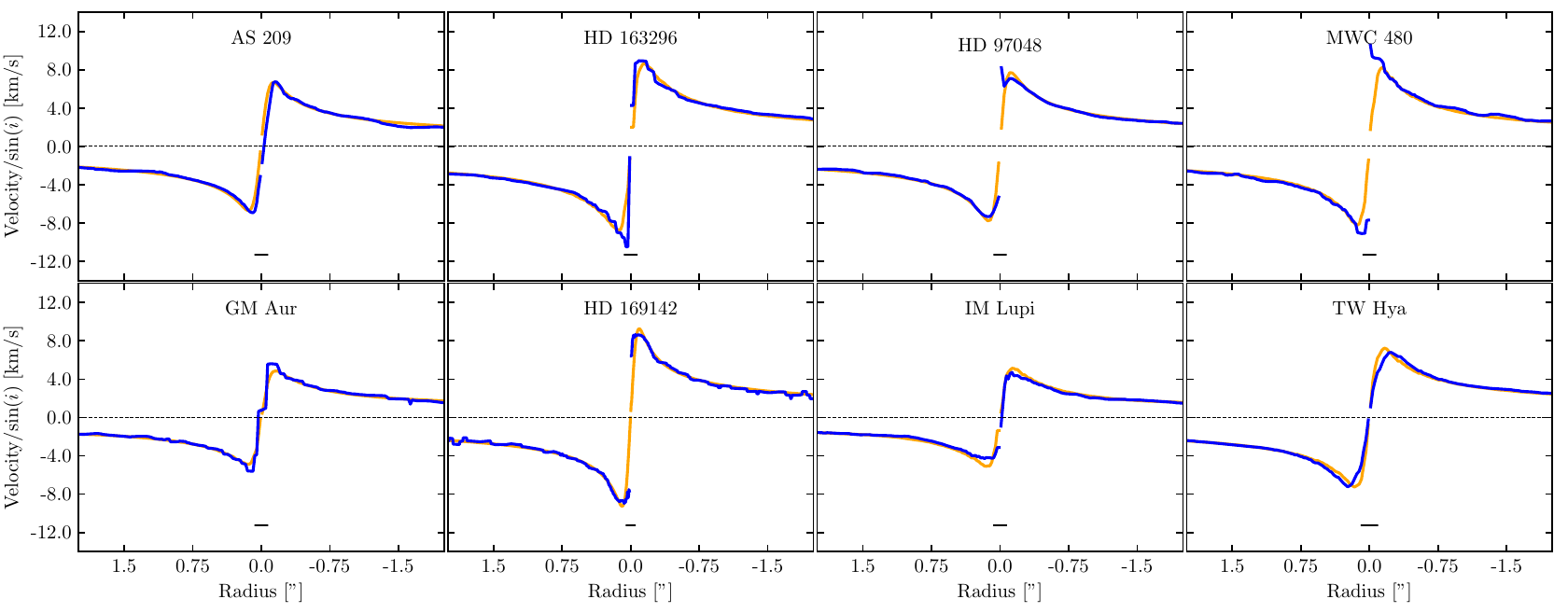}
    \caption{The rotation curve along the disc major axis of the first 8 observations listed in Table \ref{tab:obs} (blue lines) with the expected Keplerian rotation of the unperturbed disc (orange curve). For the HD 163296, GM Aur, IM Lupi, and MWC 480, we have taken the elevated emission surface into account in the Keplerian rotation model. The major axis of the beam is shown as the horizontal black line. Most of the discs in this Figure shown few deviations away from the expected Keplerian rotation. Where such deviations do exist, they tend to be within one or two beam sizes from the centre of the disc, where the rotation profile of the disc is unresolved. IM~Lupi shows some larger deviations, however this is likely due to uncertainties in modelling the highly elevated emitting layer, and it's observational effects. TW~Hya shows modest deviations away from the Keplerian model that appear several times larger than the beam, but these may be due to the dust continuum subtraction process. }
    \label{fig:rot_curve_obs1}
\end{figure*}

\begin{figure*}
    \centering
    \includegraphics[width=1.0\linewidth]{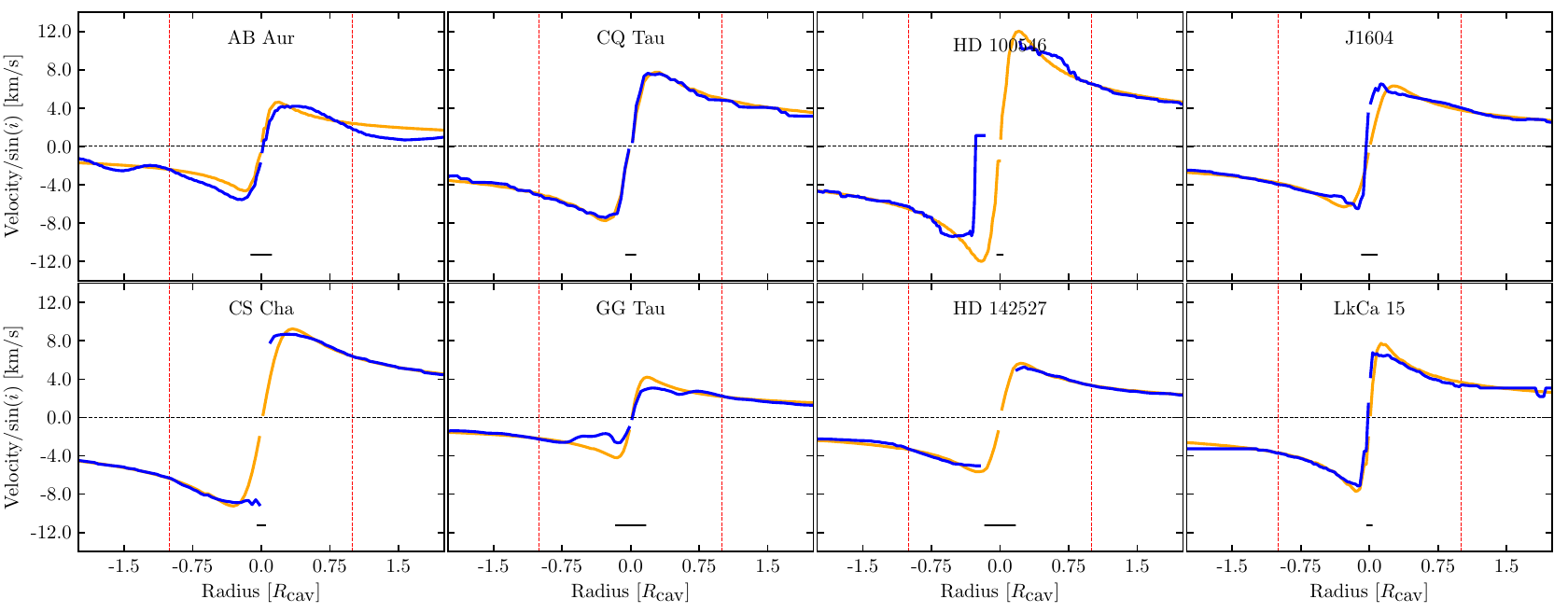}
    \caption{As in Figure \ref{fig:rot_curve_obs1} but for the last 8 discs listed in Table \ref{tab:obs}. CS Cha, GG Tau, and HD~142527 are confirmed circumbinary(-triple) discs. While all three of these discs so show some perturbations away from Keplerian, they are most obvious in GG~Tau. The optically thin $^{13}$CO and $^{12}$CO cavities inside of HD~142527 and CS~Cha limit our ability to spatially resolve their gas kinematics. Modest perturbations are also seen inside the cavities of J1604 and LkCa~15. }
    \label{fig:rot_curve_obs2}
\end{figure*}

We show the velocity close to the disc major axis of the first 8 discs of Table \ref{tab:obs} in Figure \ref{fig:rot_curve_obs1}, and the last 8 discs of Table \ref{tab:obs} in Figure \ref{fig:rot_curve_obs2}. The observed velocities are taken assuming either an elevated emission surface (for GM Aur, HD~163296, IM~Lupi, and MWC~480, where the emission surfaces are taken from \cite{mapsiv}) or pure Keplerian rotation (all other observations). We fit each velocity map with {\sc eddy} and use the resulting best-fit Keplerian model 2D velocity map to obtain the velocities along the same path close to the disc major axis as the observations. 
In both Figures we scale the line-of-sight velocity by $\sin (i)$, where $i$ is the inclination of the system. The observations are shown with the blue lines, while the models are shown with the orange lines.

Starting with Figure \ref{fig:rot_curve_obs1}, we notice that close to the central regions of the disc there can be large deviations from the best-fit model. Smearing of the high velocity components close to the central regions, along with uncertainties in the height of the emitting layer, can cause these effects. We note that these perturbations tend to be about one or two beam sizes in extent. Larger-scale perturbations are seen in IM~Lupi and TW~Hya. In IM~Lupi, the deviations are likely originating due to the CO emitting layer and disc inclination producing a complex CO emission line profile which not taken into account in our model. TW~Hya appears to have kinematic perturbations that are several beam widths across which may be due to continuum subtraction process \citep[e.g. see][]{boehler2021}.

Turning our attention to Figure \ref{fig:rot_curve_obs2}, we note that the $x$-axis has been rescaled to the size of the cavity for each of the discs, as listed in Table \ref{tab:obs}. Several discs presented are confirmed circumbinary(or -triple) discs. These are GG Tau \citep{leinert1993}, HD~142527 \citep{biller2012}, and CS~Cha \citep{guenther2007, nguyen2012}. Starting with GG~Tau, CO emission is seen within the cavity region, allowing for the highly perturbed kinematics to be traced \citep{phuong2020}. Significant deviations from the best-fit Keplerian rotation curve are seen, which are consistent with the perturbations we see in our models in Figure \ref{fig:rot_curve}. Deviations from Keplerian rotation are also seen in the rotation profiles of CS~Cha and HD~142527, but since CO emission is missing inside of the cavity, larger perturbations are not seen. 

Two other discs stand out as having large perturbations inside of their cavities: AB Aur and HD~100546. AB Aur has a claimed detection of a companion by \cite{currie2022}, and their astrometric fitting suggests it is highly eccentric and inclined with respect to the disc. Such a body could lead to significant perturbations inside the cavity, larger than those suggested by our planet-hosting disc models in Paper I. However it is not clear whether this detection is a feature on the disc surface or truly an embedded protoplanet \citep[][however see \citealt{currie2024}]{zhou2023, biddle2024}. The outer disc appears to be highly perturbed, and potentially gravitationally unstable (Speedie et al. in press), which may be responsible for the perturbations in this region. But the central cavity is optically thin in $^{13}$CO \citep{riviere2020}, possibly ruling out GI as the primary source of perturbations in this region.

HD~100546 has had several claimed detections of protoplanets, \rr{both inside \citep{currie2015} and outside \citep{quanz2013} the central cavity}. Figure \ref{fig:rot_curve} shows that planetary mass companions do not produce large amplitude perturbations inside of the cavities they generate. The highly perturbed kinematics inside of the cavity are suggestive of a more massive companion generating the central cavity and the other observed morphological features seen in this disc. 

\begin{figure*}
    \centering
    \includegraphics[width=\linewidth]{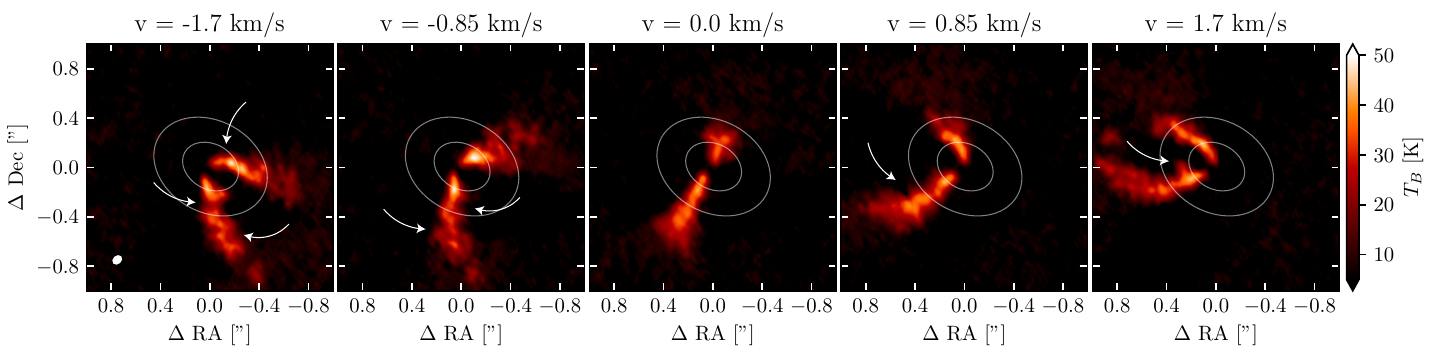}
    \caption{The CO (3-2) emission around LkCa~15. The white ellipses mark the inner and outer extent of the two bright continuum emission rings \protect{\citep{facchini2020}}. Kinks in the iso-velocity curves are seen inside and in proximity to the continuum emission, and are marked with the white arrows.}
    \label{fig:lkca15}
\end{figure*}

Modest perturbations are also seen in both J1604 and LkCa~15. For J1604, it is not entirely clear whether these perturbations are rotational or out of the plane of the disc \citep{stadler2023}. It is feasible that either a planetary mass companion on a non-circular orbit, or a more massive companion, can produce such perturbations (see Figure \ref{fig:dens_res_irs}). In LkCa~15, the relatively low signal-to-noise, sparse spectral resolution, and relatively large disc flaring, make interpreting the perturbed kinematics difficult. However the channel maps, shown in Figure \ref{fig:lkca15}, reveal that there is a large velocity kink in proximity to the cavity. In \citetalias{calcino2023}, we showed that such a signature is produced by a stellar mass companion inside the cavity. Deeper and higher spectral resolution observations will help to resolve the nature of these perturbations

\section{Summary}\label{sec:sum}

Circumbinary discs generate diverse kinematic and morphological features, some of which are observed in transitional discs that are not yet known to be circumbinary. By using simulations and synthetic observations, we have shown that Doppler-flips, spiral arms, and vortex-like features can appear in the velocity residuals and potentially be misinterpreted as evidence for planet-disc interactions. Two of these features have previously been reported in transition discs. Firstly, vortex-like kinematics were reported by \cite{boehler2021} spatially co-located with the dust trap in HD~142527. We show in Section \ref{sec:hd14} that the vortex-like kinematics can be reproduced in a circumbinary disc model that does not contain a vortex. The complexity of the kinematics of circumbinary discs mean that it will be difficult to be confidently attribute velocity perturbations inside of the dust trap to vortical motion. Secondly, a Doppler-flip in the disc around HD~100546 was reported by \cite{perez2020} and attributed to a multi-Jupiter mass planet. Recently, \cite{casassus2022} attribute the Doppler-flip to a mass eruption event from a $\sim 10 $ M$\textrm{E}$ planet. In Section \ref{sec:hd100} we show that an eccentric internal companion can reproduce the observed Doppler-flip, as well as other significant velocity perturbations around the central cavity. 

Additionally, we analyse the rotation curves of 16 protoplanetary and transitional discs in the literature. We show that two transition discs, AB Aur and HD~100546, have highly perturbed rotation curves inside of their cavities, which may be consistent with our proposition that such perturbations are generated by massive (i.e. above planetary mass) companions. Other transition discs, such as J1604 and LkCa~15 have peculiar kinematic signatures inside of their cavities which appear consistent with companion-disc interactions. 

The gas kinematics of transition discs can reveal what physical mechanism is responsible for their origin. A sizeable and high fidelity sample, such as that soon to be released by the {\sc exoALMA}\footnote{\url{https://www.exoalma.com/}} survey, will allow for a more detailed study on the statistical properties of the kinematic perturbations seen inside the cavities of transition discs.

\section*{Acknowledgements}
We thank the anonymous referee for their useful comments which helped improved the clarity of the manuscript. This work is supported by the National Natural Science Foundation of China under grant No. 12233004 and 12250610189.
Je.S. is supported by the Natural Sciences and Engineering Research Council of Canada (NSERC).
DJP and CP are grateful for Australian Research Council funding via DP180104235, DP220103767 and DP240103290.
Jo.S. has received funding from the European Research Council (ERC) under the European Union’s Horizon 2020 research and innovation programme (PROTOPLANETS, grant agreement No. 101002188).
This paper makes use ALMA data with project IDs listed in Table \ref{tab:obs}. ALMA is a partnership of ESO (representing its member states), NSF (USA) and NINS (Japan), together with NRC (Canada), MOST and ASIAA (Taiwan), and KASI (Republic of Korea), in cooperation with the Republic of Chile. The Joint ALMA Observatory is operated by ESO, AUI/NRAO and NAOJ.

\section*{Data Availability Statement}
The simulation dump files and selected radiative transfer models will be made available at \url{10.5281/zenodo.10706132}. 
Observational data from MAPS can be obtained from \url{https://alma-maps.info/data.html}. 
Additional observational data can be found in the citations in the caption of Table 2, or by contacting the authors. {\sc phantom} is publicly available at \url{https://github.com/danieljprice/phantom}, while {\sc mcfost} is available at \url{https://github.com/cpinte/mcfost}.




\bibliographystyle{mnras}
\bibliography{paper} 





\bsp	
\label{lastpage}
\end{document}